\documentclass[aps,prl,reprint,superscriptaddress]{revtex4-1}
\usepackage{graphicx}
\usepackage{dcolumn}
\usepackage{bm}
\usepackage{amsmath}

\usepackage{amsfonts}
\usepackage{textcomp} 

\usepackage{verbatim}
\usepackage {ulem}

\usepackage{color}
\usepackage{array}
\usepackage{makecell}

\newcolumntype{C}{>{\centering\arraybackslash$}m{2cm}<{$}}

\begin{document}

\title{Low damping magnetic properties and perpendicular magnetic anisotropy with strong volume contribution in the Heusler alloy Fe$_{1.5}$CoGe }

\author{Andres Conca}

\email{conca@physik.uni-kl.de}

\affiliation{Fachbereich Physik and Landesforschungszentrum OPTIMAS, Technische Universit\"at
Kaiserslautern, 67663 Kaiserslautern, Germany}

\author{Alessia Niesen}

\author{G\"unter Reiss}

\affiliation{Center for Spintronic Materials and Devices, Physics Department, Bielefeld University, 100131 Bielefeld, Germany}

\author{Burkard Hillebrands}

\affiliation{Fachbereich Physik and Landesforschungszentrum OPTIMAS, Technische Universit\"at
Kaiserslautern, 67663 Kaiserslautern, Germany}

\date{\today}

\begin{abstract}
We present a study of the dynamic magnetic properties of TiN-buffered epitaxial thin films of the Heusler alloy Fe$_{1.5}$CoGe. Thickness series annealed at different temperatures are prepared and the magnetic damping is measured, a lowest value of $\alpha=2.18\times 10^{-3}$ is obtained. The perpendicular magnetic anisotropy properties in Fe$_{1.5}$CoGe/MgO are also characterized. The evolution of the interfacial perpendicular anisotropy constant $K^{\perp}_{\rm S}$ with the annealing temperature is shown and compared with the widely used CoFeB/MgO interface. A large volume contribution to the perpendicular anisotropy of $(4.3\pm0.5)\times 10^{5}$ $\rm J/m^3$ is also found, in contrast with vanishing bulk contribution in common Co- and Fe-based Heusler alloys.

\end{abstract}

\maketitle


The need for strong perpendicular magnetic anisotropy (PMA) \cite{tin2017,takamura-co2fesi,kamada-cfms,lufbrook-cfms-pma,lufbrook-cmnga-pma} and low damping properties \cite{oogane-cfs-damping,cfa2018,mizukami-cfa-damping,sterwerf2016} in next-generation spin-transfer-torque magnetic memory (STT-MRAM) generates a large interest towards Heusler alloys.
 In addition, large tunneling magnetoresistance (TMR) ratios with MgO tunneling barriers have been reported for several of them \cite{ando-co2mnsi,inomata-cfas}.  For the application in devices based on spin transfer torque switching, a low damping parameter $\alpha$ is important since the critical switching current is proportional to $\alpha M_{\rm S}^2$ \cite{diao2007} for in-plane magnetized films. With perpendicular magnetization, the critical current is further reduced and it is proportional to $\alpha M_{\rm S}$ \cite{wang2013}. Therefore, a large effort is directed to study the PMA properties of  Heusler alloys with low damping. 

\begin{figure}[t]
    \includegraphics[width=0.9\columnwidth]{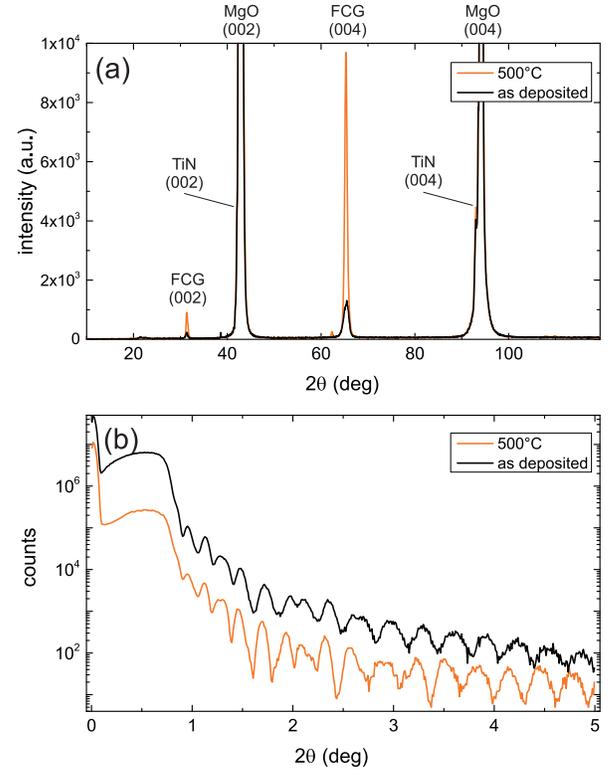}
	  \caption{\label{xrd}(Color online) (a) X-ray diffraction patterns of 40~nm thin Fe$_{1.5}$CoGe layers as-deposited, and annealed at 500$^{\circ}$C. The (002) superlattice and the fundamental (004) peak of the Fe$_{1.5}$CoGe are clearly visible, confirming the partial B2 crystalline order. (b) X-ray reflectometry data corresponding to the samples in (a).}
\end{figure}

For the PMA  of thin Heusler films, the interface-induced perpendicular anisotropy is essential and its strength is given by the perpendicular interfacial anisotropy constant $K^{\perp}_{\rm S}$. The interface properties, and therefore the value of $K^{\perp}_{\rm S}$, are strongly influenced by the conditions of the annealing, which is required to improve the crystalline order of the Heusler and MgO layers  and to achieve large TMR values.

Here, we report on the evolution of the PMA properties with annealing of the PMA properties  in Fe$_{1.5}$CoGe with a MgO interface, by measuring different thickness series and a comparison is made with the well-known CoFeB/MgO interface. The Gilbert damping parameter $\alpha$ change with varying thickness and annealing temperature is also discussed.


The films were grown by sputtering, Rf-sputtering was used for the MgO deposition and
dc-sputtering for the rest. For Fe$_{1.5}$CoGe, the layer stack is MgO(S) / TiN(30) / Fe$_{1.5}$CoGe($d$) / MgO(7) / Si(2) with $d=80$, 40, 20, 15, 11 and 9~nm. Four series were deposited and three of them annealed for one hour at 320$^{\circ}$C, 400$^{\circ}$C and 500$^{\circ}$C.  For CoFeB, the layer stack structure is MgO(s) / Ta(5) / Ru(30) / Ta(10) / MgO(7) / CoFeB($d$) / MgO(7) / Ta(5) / Ru(2) $d=80$, 40, 20, 15, 11, 9, 7 and 5~nm. The annealing was performed at 325$^{\circ}$C and 360$^{\circ}$C.

The dynamic properties and material parameters were studied by measuring the ferromagnetic resonance using
a strip-line vector network analyzer (VNA-FMR). For this, the samples were placed facing the strip-line and
the S$_{12}$ transmission parameter was recorded.

Crystallographic properties of the CFA thin films were determined using x-ray diffraction (XRD) measurements in a Philips X'Pert Pro diffractometer equipped with a Cu anode.


The XRD data corresponding to two 40~nm thick samples in the as-deposited state and annealed at 500$^{\circ}$C are shown in Fig.~\ref{xrd}(a). The (002) superlattice and the fundamental (004) peak of Fe$_{1.5}$CoGe can be observed already for the as-deposited state but they experience a strong intensity increase with the thermal treatment. The TiN layer acts as a seed layer  and its role in improving growth has been reported also for other alloys \cite{tin2015,tin2017}. Due to the similar lattice constant of TiN and MgO, the TiN film diffraction peaks are close to the substrate reflections and therefore difficult to separate. The films are B2-ordered, the presence of the (111) is not proven and therefore L2$_1$ order cannot be confirmed.  

\begin{figure}[t]
    \includegraphics[width=0.95\columnwidth]{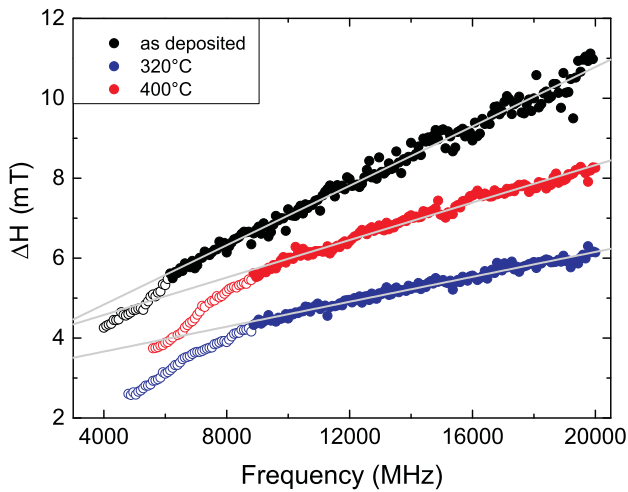}
	  \caption{\label{linewidth}(Color online) Linewidth dependence on the frequency for Fe$_{1.5}$CoGe thin films with a thickness of 20~nm for different annealing temperatures. The data sets have a vertical offset to improve visibility. The lines correspond to a linear fit to extract the damping parameter $\alpha$. The hollow points are not considered for the fits (see main text).}
\end{figure}

Figure~\ref{xrd}(b) shows X-ray reflectometry (XRR) data for the same films as in the (a) panel. The large number of oscillations prove the low roughness of the interfaces. This is due to the low roughness below 1~nm of the TiN buffer \cite{tin2015}. The similarity between both data sets also proves that the topology of the interfaces do not vary in the studied temperature range.


Figure \ref{linewidth} shows the dependence of the field linewidth $\Delta H$ of the FMR peak on the resonance frequency $f_{\rm FMR}$ for the sample series  with no thermal treatment (as-deposited) and for the series annealed at 320$^{\circ}$C and 400$^{\circ}$C. In order to prevent  poor visibility due to data overlapp, the sets are shifted in the vertical axis, except the one corresponding to  320$^{\circ}$C. The actual linewidth  at 6~GHz is in the range $3.25\pm0.15$~mT. The lines represent the result to a linear fit to Eq.~\ref{alphaeq} to extract the damping parameter $\alpha$:

\begin{equation} \label{alphaeq}
\mu_0\Delta H= \mu_0\Delta H_0 + \frac{4 \pi \alpha f_{\rm FMR}}{\gamma}.
\end{equation}
Here, $\Delta H_0$ is the inhomogenous broadening and is related to film quality, and $\gamma$ is the gyromagnetic ratio.

 A deviation from this simple linear behavior is observed for the lower frequency range and these points have not been considered for the fit (hollow circles). This faster increase of linewidth with frequency is common in fully epitaxial Heusler layers \cite{oogane2016,mizukami-cfa-damping} and has been related with an increased anisotropic two-magnon scattering in the thin films for low frequency values resulting in an  anisotropic $\Delta H$. This anisotropy is not exclusive to Heusler alloys but it is expected in any epitaxial ferromagnetic film  \cite{woltersdorf,aria-mills,zakeri}. The exact conditions for observation, however, depend on the material parameters and the spin-wave dispersion.  For instance, in epitaxial Fe films, the low frequency $\Delta H$ behavior deviates only from a linear behavior when magnetic dragging due to crystalline anisotropy is dominant \cite{2ms2018}.
 
\begin{figure}[t]
    \includegraphics[width=0.95\columnwidth]{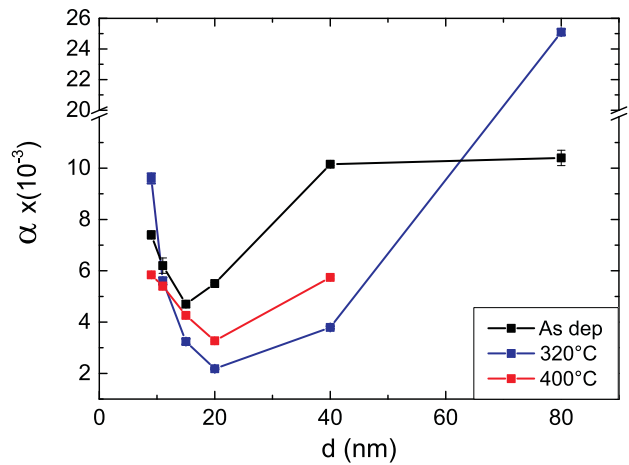}
	  \caption{\label{alpha}(Color online) Dependence of the Gilbert damping parameter $\alpha$ on the thickness $d$ for three sample series: as-deposited, annealed at 320$^{\circ}$C, and annealed at 400$^{\circ}$C. }
\end{figure}

An additional sample set annealed at 500$^{\circ}$C showed no visible FMR peak pointing to a degradation of the magnetic properties of Fe$_{1.5}$CoGe for high annealing temperature. This is in constrast with  Co-based Heusler alloys where large temperatures are typically required for optimal properties. For instance, for Co$_2$FeAl, lowest damping is achieved at 600$^{\circ}$C \cite{mizukami-cfa-damping} and for Co$_2$MnSi, very low damping is still present at 750$^{\circ}$C \cite{andrieu-cms}.

\begin{figure*}[t]
    \includegraphics[width=1.0\textwidth]{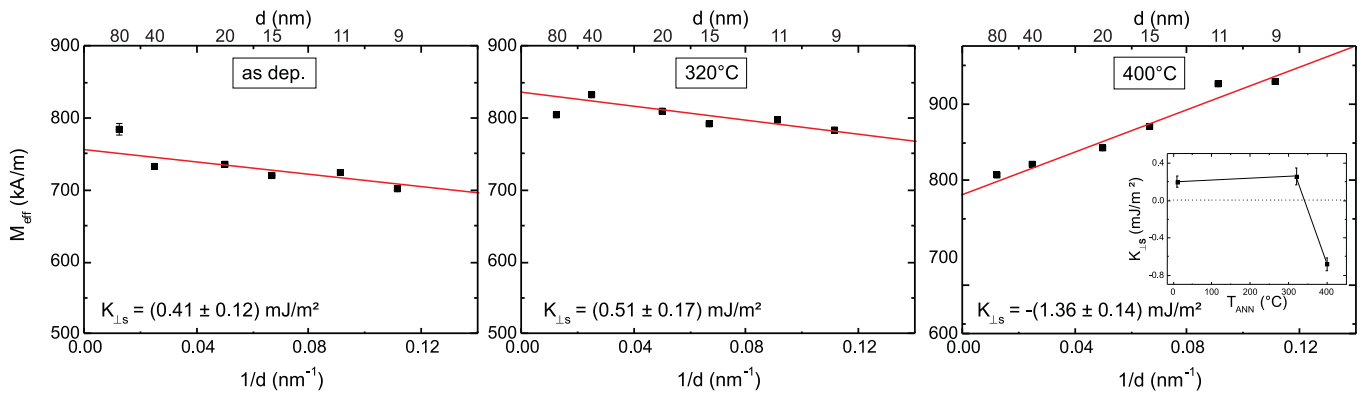}
	  \caption{\label{ks}(Color online) Dependence of $M_{\rm eff}$ extracted from the Kittel fit on the inverse thickness $1/d$ for three Fe$_{1.5}$CoGe sample series:  as-deposited,  annealed at 320$^{\circ}$C and annealed at 400$^{\circ}$C. The lines are a fit to Equation~\ref{kp}. The inset shows the evolution of $K^{\perp}_{\rm S}$ with the annealing temperature.}
\end{figure*}

The results for the damping parameter $\alpha$ obtained from the linear fits are summarized in Fig.~\ref{alpha}. A reduction of damping is observed when comparing the as-deposited samples to the annealed ones but the samples annealed at 400$^{\circ}$C show larger damping than the ones annealed at 320$^{\circ}$C. Combined with the absence of an FMR peak for 500$^{\circ}$C, this reinforces  the conclusion that the optimal annealing temperature for good dynamic properties of Fe$_{1.5}$CoGe is low. The lowest damping  is $2.18 \pm 0.03 \times 10^{-3}$ for the 20~nm thick film annealed at 320$^{\circ}$C. For Co-based Heusler alloys, the lowest reported damping  is achieved to be $7 \times 10^{-4}$ in Co$_2$MnSi \cite{andrieu-cms}. For Co$_2$FeAl, values around 1-3$\times 10^{-3}$ are reported depending on the annealing conditions \cite{mizukami-cfa-damping,cfa2018}. Concerning Fe-based alloys, values in the range of 1.2-1.9$\times 10^{-3}$ are reported for Fe$_{1+x}$Co$_{2-x}$Si \cite{sterwerf2016}, for Fe$_{2}$Cr$_{1-x}$Co$_{x}$Si  $\alpha$  varies between 9$\times 10^{-3}$ with the lowest value of 8$\times 10^{-4}$ for  Fe$_{2}$CoSi \cite{he2017}. Therefore, the obtained value for the damping parameter in our alloy is in the lower range of previously reported ones and slighty reduced compared to those reported for the related alloy CoFeGe \cite{lee2009}. It is also smaller than the ones reported for widely used polycrystalline CoFeB \cite{bilzer,cofeb-annealing} and permalloy \cite{ruiz2015,luo2014,py2007,kraut2018}.

The thickness dependence of $\alpha$ shows a minimum around 20~nm and an increase for larger and smaller thicknesses. This behavior has been already observed for Co$_2$FeAl \cite{cfa2018}, and the reasons are  similar to that alloy and different for the two thickness ranges. For soft magnetic thin films, a strong damping increase  with increasing thickness is expected starting at a certain value. An example can be found for NiFe in the literature \cite{chen}.  The reason  is a non-homogeneous magnetization state for thicker films which opens new loss channels via increased magnon scattering. For the thinner films, the damping increase is due to two reasons. When the thickness is reduced and the effect of the interface anisotropy is becoming larger the magnetization state is becoming more inhomogeneous due to the counterplay between the demagnetization field and the anisotropy field \cite{usov}. In addition, other effects related to an increased role of surface roughness with decreasing film thickness play also a role. 


 The effective magnetization $M_{\rm eff}$ is extracted using a fit to Kittel's formula \cite{kittel} to the dependence of the resonance field $H_{\rm FMR}$ on the resonance frequency $f_{\rm FMR}$. 
For a more detailed description of the FMR measurement and analysis procedure see Ref.~\cite{fept}. $M_{\rm eff}$ is  related to the saturation magnetization of the film by \cite{vries1996,beaujour,belme2} 

\begin{equation} \label{kp}
M_{\rm eff}= M_{\rm s}-H^{\perp}_{K}= M_{\rm s}-\frac{1}{\mu_0 M_{\rm s}}\left(\frac{K^{\perp}_{\rm S}}{d}+K^{\perp}_{\rm V}\right)
\end{equation}
where $K^{\perp}_{\rm S}$ and $K^{\perp}_{\rm V}$ are the perpendicular surface (or interfacial) and the bulk anisotropy constants, respectively.

Figure~\ref{ks} shows the dependence of M$_{\rm eff}$  on the inverse thickness $1/d$ for the three sample series: as-deposited,  annealed at 320$^{\circ}$C and annealed at 400$^{\circ}$C. The slope provides the value of  $K^{\perp}_{\rm S}$. The constant shows a positive value for the as-deposited series, $0.41\pm0.12$~mJ/m$^2$, i.e. favouring a perpendicular orientation of the magnetization. However, the value is small in absolute value and it grows only slighthy upto $0.51\pm0.17$~mJ/m$^2$ when the samples are treated at 320$^{\circ}$C. The annealing at 400$^{\circ}$C changes the situation drastically. The value of $K^{\perp}_{\rm S}$ is much larger, $-1.36\pm0.14$~mJ/m$^2$,  but it also suffers a change of sign which implies that the interface induces an in-plane orientation of the magnetization. It is remarkable that this change of the magnetic properties of the interface developes without a large modification of the morphology, as proven by the XRR data shown in Fig.~\ref{xrd}(b). The inset in Fig.~\ref{ks} summarizes the dependence of $K^{\perp}_{\rm S}$ on the annealing temperature.

Taking into account the saturation magnetization $M_{\rm s}$ obtained by alternating gradient magnetometer (AGM), $1100\pm120$~kA/m, it is possible to determine also the volume contribution to the perpendicular anisotropy to be  $(4.3\pm0.5)\times 10^{5}\frac{\rm J}{\rm m^3}$.  This large value ensures, that for a 1.5~nm thin film and even for the 320$^{\circ}$C case, the PMA properties are dominated by the bulk contribution. Recently, we reported on the evolution of the PMA properties of the Co$_2$FeAl/MgO interface \cite{cfa2018} and the situation is very different for that alloy. First, no interface-generated PMA is present in the as-deposited samples and it only appears after annealing. Second, $K^{\perp}_{\rm S}$ is always positive and larger than for Fe$_{1.5}$CoGe/MgO and  the absence of a remarkable volume contribution makes the PMA there controlled only by the interface. Also in the related alloy Co$_{20}$Fe$_{50}$Ge$_{30}$ there is no bulk contribution to the perpendicular anisotropy and a interface contribution ($0.9$~mJ/m$^2$) larger than that obtained here \cite{dinga}. The most probable reason for the difference is the lower Fe content in our case. The presence of this strong bulk contribution to PMA is quite remarkable since it is absent in common Co- and Fe-based Heusler alloys and only observed in tetragonally distorted MnGa or MnGe related Heusler alloys \cite{mnfega2018,mnfege2016,mnga2017}.

\begin{figure}[b]
    \includegraphics[width=0.8\columnwidth]{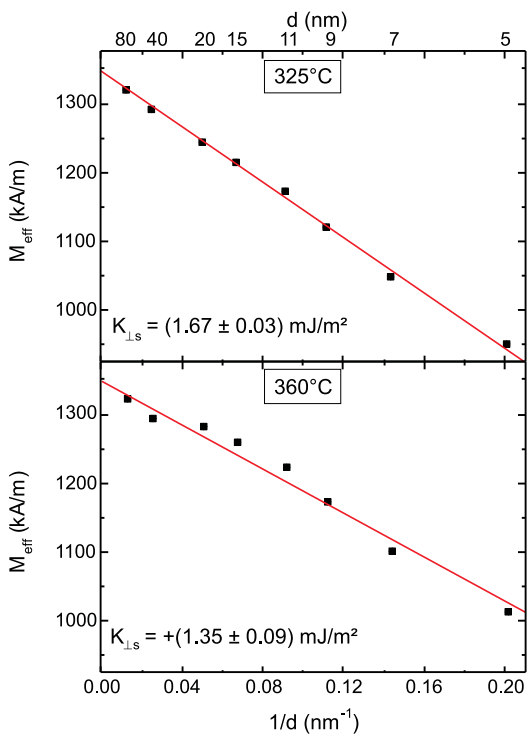}
	  \caption{\label{cofeb}(Color online) Dependence of $M_{\rm eff}$ extracted from the Kittel fit on the inverse thickness $1/d$ for two MgO/CoFeB/MgO sample series:  annealed at 325$^{\circ}$C and annealed at 360$^{\circ}$C. The lines are a fit to Equation~\ref{kp} with a prefactor 2 (see text). }
\end{figure}

For comparison, the PMA properties of the widely used CoFeB/MgO interface were also measured. The data is shown in Fig.~\ref{cofeb} for annealing temperatures of 325$^{\circ}$C and 360$^{\circ}$C. An as-deposited series was not characterized since CoFeB is amorphous in that state. The lines are a fit to Eq.~\ref{kp} with a prefactor 2 before $K^{\perp}_{\rm S}$ to account for the presence of two interfaces since a trilayer system MgO/CoFeB/MgO was used.

The CoFeB/MgO interface shows a robust interface perpendicular anisotropy, three times larger than for Fe$_{1.5}$CoGe/MgO,  and slighty decreases with temperature. The bulk contribution is zero or too small to be detectable. These results are comparable to the literature \cite{worledge2011, liu2011, liu2014}. From this we conclude that while the Co$_2$FeAl \cite{cfa2018} and CoFeB/MgO are very similar in absolute values, thermal evolution and relative weight of interface and bulk contribution to PMA,  Fe$_{1.5}$CoGe/MgO differs strongly.

A comment must be made concerning the relation between the values of $K^{\perp}_{\rm S}$ shown in Fig.~\ref{ks} and the increase in damping for the thinnest films shown in Fig.~\ref{alpha}. When discussing the increase of $\alpha$, a possible connection is suggested with the interplay between  the demagnetization field and the perpendicular anisotropy field. An additional hint supporting this interpretation is given by the fact that the for the 400$^{\circ}$C samples series, for which no interface PMA is present, the relative increase in $\alpha$ from the series minimum is the smallest, as well as the absolute value for the thinnest
 film.

 In summary, the damping properties of the Heusler alloy Fe$_{1.5}$CoGe and the perpendicular magnetic anisotropy of the  Fe$_{1.5}$CoGe/MgO  system have been studied. From the  thickness dependent magnetic properties for as-deposited and annealed series we obtained a minimum value for $\alpha$ of $2.18 \pm 0.03 \times 10^{-3}$ for a 20~nm thick film. The evolution of the interface perpendicular anisotropy constant on the annealing temperature is shown and compared with the standard interface CoFeB/MgO.  We found a large and dominant volume contribution to the PMA, which differs from CoFeB or other well studied alloys as Co$_2$FeAl and the interface contribution suffers a sign change depending on the annealing temperature. We explained the increase on damping with decreasing thickness in terms of a counterplay between demagnetizing field and interface PMA and correlate it with the obtained values for $K^{\perp}_{\rm S}$.
 


\section*{Acknowledgements}

Financial support by  M-era.Net through the HEUMEM project is gratefully acknowledged.


\begin{thebibliography}{sotief}

\bibitem{tin2017} A.~Niesen, J.~Ludwig, M.~Glas, R.~Silber, J.-M.~Schmalhorst, E.~Arenholz, and G.~Reiss,  J.~Appl.~Phys. \textbf{121}, 223902  (2017).

\bibitem{takamura-co2fesi}Y.~Takamura, T.~Suzuki, Y.~Fujino, and S.~Nakagawa, J.~Appl.~Phys. \textbf{115}, 17C732 (2014). 

\bibitem{kamada-cfms}T.~Kamada, T.~Kubota, S.~Takahashi, Y.~Sonobe, and K.~Takanashi, IEEE Trans. on Magn. \textbf{50}, 2600304 (2014).

\bibitem{lufbrook-cfms-pma}B.~M.~Ludbrook, B.~J.~Ruck, and S.~Granville, J.~Appl.~Phys. \textbf{120}, 013905 (2016).

\bibitem{lufbrook-cmnga-pma}B.~M.~Ludbrook, B.~J.~Ruck, and S.~Granville, Appl.~Phys.~Lett. \textbf{110}, 062408 (2017).

\bibitem{oogane-cfs-damping} M.~Oogane, R.~Yilgin, M.~Shinano, S.~Yakata, Y.~Sakuraba, Y.~Ando, and T.~Miyazaki, J.~Appl.~Phys. \textbf{101}, 09J501 (2007).

\bibitem{mizukami-cfa-damping}S.~Mizukami, D.~Watanabe, M.~Oogane, Y.~Ando, Y.~Miura, M.~Shirai, and T.~Miyazaki, J.~Appl.~Phys. \textbf{105}, 07D306 (2009).

\bibitem{cfa2018}A.~Conca, A.~Niesen, G.~Reiss, and B.~Hillebrands,  J.~Phys.~D: Appl.~Phys. \textbf{51}, 165303 (2018).

\bibitem{sterwerf2016}C.~Sterwerf, S.~Paul, B.~Khodadadi, M.~Meinert, J-M.~Schmalhorst, M.~Buchmeier, C.~K.~A. Mewes, T.~Mewes, and G.~Reiss, J.~Appl.~Phys. \textbf{120}, 083904 (2016).



\bibitem{inomata-cfas}N.~Tezuka, N.~Ikeda, S.~Sugimoto, and K.~Inomata, Jpn. J.~Appl.~Phys. \textbf{46}, L454 (2007).

\bibitem{ando-co2mnsi}S.~Tsunegi, Y.~Sakuraba, M.~Oogane, K.~Takanashi, and Y.~Ando, Appl.~Phys.~Lett. \textbf{93}, 112506 (2008).

\bibitem{diao2007}Z.~Diao, Z.~Li, S.~Wang, Y.~Ding, A.~Panchula, E.~Chen, L.-C.~Wang, and Y.~Huai, J.~Phys.~D: Appl.~Phys. \textbf{19}, 165209 (2007).

\bibitem{wang2013}K.~L.~Wang, J.~G.~Alzate, and P.~K.~Amiri, J.~Phys.~D: Appl.~Phys. \textbf{46}, 074003 (2013).

\bibitem{tin2015}A.~Niesen, M.~Glas, J.~Ludwig, J.-M.~Schmalhorst, R.~Sahoo, D.~Ebke, E.~Arenholz, and G.~Reiss, J.~Appl.~Phys. \textbf{118}, 243904 (2015).






\bibitem{oogane2016}  M.~Oogane and S.~Mizukami, Phil. Trans. R. Soc. A  \textbf{369}, 3037 (2011).



\bibitem{woltersdorf} G.~Woltersdorf and B.~Heinrich, Phys.~Rev.~B \textbf{69}, 184417 (2004)

\bibitem{aria-mills} R.~Arias and D.~L.~Mills, Phys.~Rev. B \textbf{60}, 7395 (1999). 

\bibitem{zakeri}Kh.~Zakeri, J.~Lindner, I.~Barsukov, R.~Meckenstock, M.~Farle, U.~von H\"orsten, H.~Wende,  W.~Keune, J.~Rocker, S.~S.~Kalarickal, K.~Lenz, W.~Kuch, K.~Baberschke, and Z. Frait, Phys. Rev B \textbf{76}, 104416 (2007).

\bibitem{2ms2018} A.~Conca, S.~Keller, M.~R.~Schweizer, E.~Th.~Papaioannou, and B.~Hillebrands, Phys.~Rev. B \textbf{98}, 214439 (2018).


\bibitem{andrieu-cms} St\'ephane Andrieu, Amina Neggache, Thomas Hauet, Thibaut Devolder, Ali Hallal, Mairbek Chshiev, Alexandre M. Bataille, Patrick Le F\`evre, and Fran\c{c}ois Bertran, Phys.~Rev.~B {\bf 93}, 094417 (2016).






\bibitem{he2017}S.~He, Y.~Liu, Y.~Zheng, Q.~Qin, Z.~Wen, Q.~Wu, Y.~Yang, Y.~Wang,
Y.~Feng, K.~Leong Teo, and C.~Panagopoulos, Phys.~Rev.~ Materials. \textbf{1}, 064401 (2017).




\bibitem{lee2009}H.~Lee, Y.-H.~A.~Wang, C.~K.~A.~Mewes, W.~H.~Butler, T.~Mewes,  S.~Maat, B.~York, M.~J.~Carey, and J.~R.~Childress, Appl.~Phys.~Lett. \textbf{95}, 082502 (2009).

\bibitem{bilzer}C.~Bilzer, T.~Devolder, J.-V.~Kim, G.~Counil, C.~Chappert, S.~Cardoso, and
P.~P.~Freitas,  J.~Appl.~Phys. \textbf{100}, 053903 (2006).

\bibitem{cofeb-annealing}A.~Conca, E.~Th.~Papaioannou, S.~Klingler, J.~Greser, T.~Sebastian, B.~Leven, J.~Lösch, and B.~Hillebrands, Appl.~Phys.~Lett. \textbf{104}, 182407 (2014).

\bibitem{ruiz2015}A.~Ruiz-Calaforra, T.~Br\"acher, V.~Lauer, P.~Pirro, B.~Heinz, M.~Geilen, A.~V.~Chumak,
A.~Conca, B.~Leven, and B.~Hillebrands, J.~Appl.~Phys. \textbf{117}, 163901 (2015)

\bibitem{luo2014}C.~Luo, Z.~Feng, Y.~Fu, W.~Zhang, P.~K.~J.~Wong, Z.~X.~Kou, Y.~Zhai, H.~F.~Ding, M.~Farle, J.~Du, and H.~R.~Zhai, Phys.~Rev.~B \textbf{89}, 184412 (2014).

\bibitem{py2007}J.~O.~Rantschler, R.~D.~McMichael, A.~Castillo, A.~J.~Shapiro, W.~F.~Egelhoff, B.~B.~Maranville, D.~Pulugurtha, A.~P.~Chen, and L.~M.~Connors, J.~Appl.~Phys. \textbf{101}, 033911 (2007).

\bibitem{kraut2018}P.~Krautscheid, R.~M.~Reeve, D.~Sch\"onke, I.~Boventer, A.~Conca, A.~V.~Chumak, B.~Hillebrands, J.~Ehrler, J.~Osten, J.~Fassbender, and M.~Kl\"aui, Phys.~Rev.~B \textbf{98}, 214406 (2018).


\bibitem{chen}Y.~Chen, D.~Hung, Y.~Yao, S.~Lee, H.~Ji, and C.~Yu,  J.~Appl.~Phys. \textbf{101}, 09C104 (2007).



\bibitem{usov} N.~A.~Usov, and O.~N.~Serebryakova,   J.~Appl.~Phys. \textbf{121}, 133905 (2017).

\bibitem{kittel} C.~Kittel,  Phys. Rev. {\bf 73}, 155 (1948).

\bibitem{fept} A.~Conca, S.~Keller, L.~Mihalceanu, T.~Kehagias, G.~P.~Dimitrakopulos, B.~Hillebrands,  and E.~Th.~Papaioannou,  Phys.~Rev.~B {\bf 93}, 134405 (2016).



\bibitem{dinga}M.~Dinga and S.~J.~Poon, Appl.~Phys.~Lett. \textbf{101}, 122408 (2012).

\bibitem{vries1996}M.~Johnson, P.~Bloemen, F.~Den Broeder, and J.~De Vries, Rep.~Prog.
Phys. \textbf{59}, 1409 (1996).

\bibitem{beaujour}J.-M.~L.~Beaujour, W.~Chen, A.~D.~Kent, and J.~Z.~Sun, J.~Appl.~Phys. \textbf{99}, 08N503  (2006).

\bibitem{belme2}M.~Belmeguenai, H.~Tuzcuoglu, M.~S.~Gabor, T.~Petrisor, Jr., C.~Tiusan, D.~Berling, F.~Zighem, T.~Chauveau, S.~M.~Ch\'erif, and P.~Moch,  Phys.~Rev.~B {\bf 87}, 184431 (2013).

\bibitem{mnfega2018}A.~Niesen, N.~Teichert, T.~Matalla-Wagner, J.~Balluf, N.~Dohmeier, M.~Glas, C.~Klewe, E.~Arenholz, J-M.~Schmalhorst, and G.~Reiss, J.~Appl.~Phys. \textbf{123}, 113901 (2018).

\bibitem{mnfege2016}A.~Niesen, C.~Sterwerf, M.~Glas, J-M.~Schmalhorst, and G.~Reiss, IEEE Trans. on Magn. 52, 2600404 (2016).

\bibitem{mnga2017}K.~Z.~Suzuki, A.~Ono, R.~Ranjbar, A.~Sugihara, and S.~Mizukami, IEEE Trans. on Magn. 53, 2101004 (2017).

\bibitem{worledge2011}D.~C.~Worledge, G.~Hu, David W.~Abraham, J.~Z.~Sun, P.~L.~Trouilloud, J.~Nowak, S.~Brown, M.~C.~Gaidis, E.~J.~O'Sullivan, and R.~P.~Robertazzi, Appl.~Phys.~Lett. \textbf{98}, 022501 (2011).

\bibitem{liu2011}X.~Liu, W.~Zhang, M.~J.~Carter, and G.~Xiao, J.~Appl.~Phys. \textbf{110}, 033910 (2011).

\bibitem{liu2014}T.~Liu, Y.~Zhang, J.~W.~Cai, and H.~Y.~Pan, Sci. Rep. \textbf{4}, 5895 (2014).





 







































\end{thebibliography}
\end{document}